\def\year{2019}\relax

\documentclass[letterpaper]{article} 
\usepackage{aaai19}  
\usepackage{times}  
\usepackage{helvet}  
\usepackage{courier}  
\usepackage{url}  
\usepackage{graphicx}  

\usepackage{booktabs} 
\usepackage{listings}
\usepackage{amssymb}
\usepackage{tabularx}
\usepackage{multirow}
\usepackage{subcaption}
\usepackage{lipsum}
\usepackage{enumitem}
\setlist{leftmargin=3.5mm}
\usepackage{mathrsfs,amsmath}
\usepackage{colortbl}
\usepackage[misc]{ifsym}
\newtheorem{definition}{Definition}
\usepackage{xcolor}

\DeclareMathOperator{\otherwise}{otherwise}

\usepackage{amsmath,amssymb}
\usepackage{booktabs}

\usepackage[ruled]{algorithm2e}

\begin{document}
	
\title{\textit{AiDroid}: When Heterogeneous Information Network Marries Deep Neural Network for Real-time Android Malware Detection}
\author{\parbox{0.75\linewidth}{\centering Yanfang Ye$^{*1}$, Shifu Hou$^{1}$, Lingwei Chen$^{1}$, Jingwei Lei$^{2}$, Wenqiang Wan$^{2}$, Jiabin Wang$^{2}$, Qi Xiong$^{2}$, Fudong Shao$^{2}$} \\
	$^{1}$Department of Computer Science and Electrical Engineering\\
	West Virginia University, Morgantown, WV, USA\\
	yanfang.ye@mail.wvu.edu, \{shhou, lgchen\}@mix.wvu.edu\\
	$^{2}$Tencent Security Lab, Tencent, Guangdong, China\\
	\{lingfonglei, johnnywan, luciferwang, keonxiong, joeyshao\}@tencent.com\\
}
\maketitle

\begin{abstract}

The explosive growth and increasing sophistication of Android malware call for new defensive techniques that are capable of protecting mobile users against novel threats. In this paper, we first extract the runtime Application Programming Interface (API) call sequences from Android apps, and then analyze higher-level semantic relations within the ecosystem to comprehensively characterize the apps. To model different types of entities (i.e., \textit{app}, \textit{API}, \textit{IMEI}, \textit{signature}, \textit{affiliation}) and the rich semantic relations among them, we then construct a structural heterogeneous information network (HIN) and present meta-path based approach to depict the relatedness over apps. To efficiently classify nodes (e.g., apps) in the constructed HIN, we propose the \textit{HinLearning} method to first obtain in-sample node embeddings and then learn representations of out-of-sample nodes without rerunning/adjusting HIN embeddings at the first attempt. Afterwards, we design a deep neural network (DNN) classifier taking the learned HIN representations as inputs for Android malware detection. A comprehensive experimental study on the large-scale real sample collections from Tencent Security Lab is performed to compare various baselines. Promising experimental results demonstrate that our developed system \textit{AiDroid} which integrates our proposed method outperforms others in real-time Android malware detection. \textit{AiDroid} has already been incorporated into Tencent Mobile Security product that serves millions of users worldwide.

\end{abstract}

\section{1. Introduction}

\begin{figure}[htbp!]
	\vspace{-0.25cm}
	\centering
	\includegraphics[width=\linewidth]{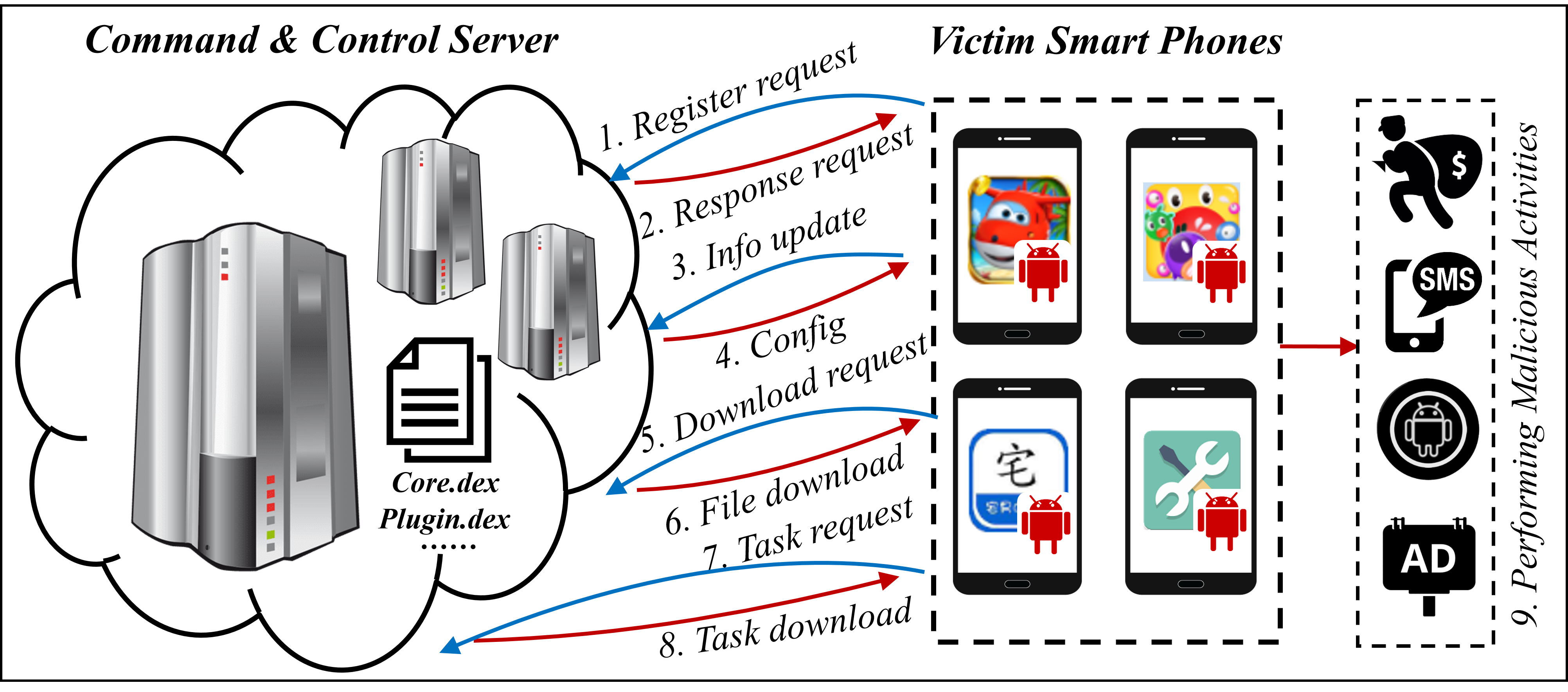}
	\vspace{-0.5cm}
	\caption{Increasingly sophisticated Android malware.} \label{fig:example}
	\vspace{-0.35cm}
\end{figure}

Due to the mobility and ever expanding capabilities, smart phones have become increasingly ubiquitous in people's everyday life performing tasks such as social networking, online banking, and entertainment. Android, as an open source and customizable operating system (OS) for smart phones, is currently dominating the smart phone market by 77.32\% \cite{Statcounter:2018}. However, due to its large market share and open source ecosystem of development, Android attracts not only the developers for producing legitimate Android applications (apps), but also attackers to disseminate malware (\textbf{\textit{mal}}icious soft\textbf{\textit{ware}}) that deliberately fulfills the harmful intent to the smart phone users (e.g., stealing user credentials, pushing unwanted apps or advertisements). Because of lacking trustworthiness review methods, developers can easily upload their Android apps including repackaged apps and malware to the official marketplace (i.e., Google Play). The presence of other third-party Android markets (e.g., Opera Mobile Store, Wandoujia) makes this problem worse. Driven by the considerable economic profits, there has been  explosive growth of Android malware which posed serious threats to the smart phone users - i.e., it's reported that there have been $4,687,008$ newly generated Android malware that infected more than $61$ million smart phones in the first half of 2018 \cite{TencentSecurityReport}. To evade the detection of mobile security products (e.g., Norton, Lookout and Tencent Mobile Security), Android malware has turned to be increasingly sophisticated. For example, as shown in Figure  \ref{fig:example}, the ``TigerEyeing'' trojan is a new kind of Command and Control (C\&C) malware that pretends to be legitimate apps (e.g., mobile games, system tools) and only executes to perform the profitable tasks on-demand. The explosive growth and increasing sophistication of Android malware call for new defensive techniques that are capable of protecting smart phone users against novel threats.

\begin{figure*}[tbp!]
\vspace{-0.1cm}
	\centering
	\includegraphics[width=\linewidth]{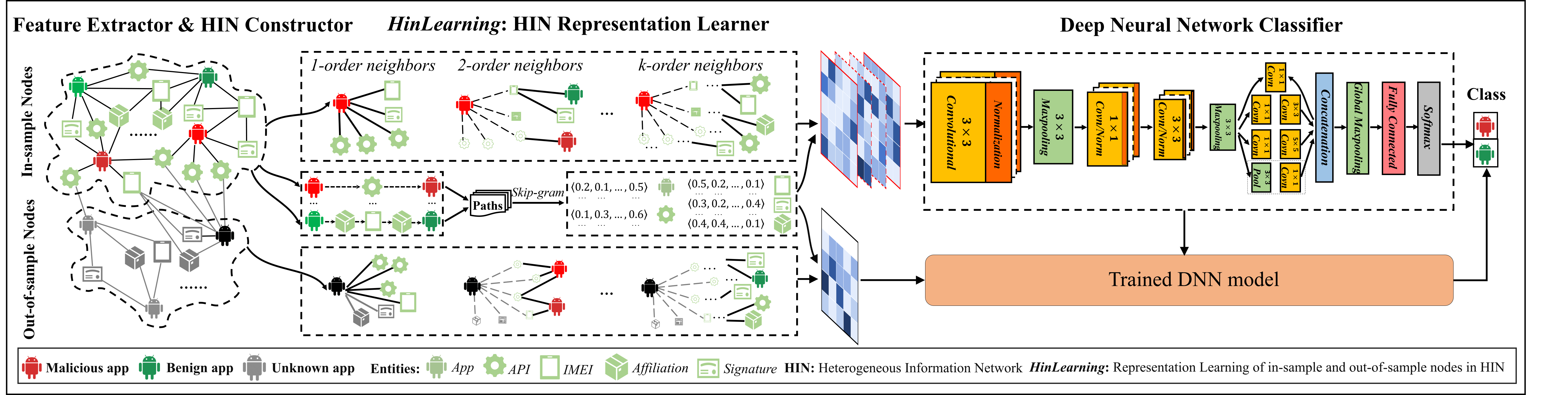}
	\vspace{-0.5cm}
	\caption{Overview of the developed system \textit{AiDroid} for real-time Android malware detection.} \label{fig:system}
	\vspace{-0.5cm}
\end{figure*}

To combat the evolving Android malware attacks, by collaboration with smart phone vendors, in this paper, we first extract the Application Programming Interface (API) call sequences from runtime executions of Android apps in users' smart phones to capture their behaviors. To comprehensively characterize Android apps, we further analyze higher-level semantic relationships within the ecosystem, such as whether two apps have similar behaviors, whether they co-exist in the same smart phone that can be identified by its unique International Mobile Equipment Identity (IMEI) number, whether they are signed by the same developer or produced by the same company (i.e., affiliation), etc. To model such complex relationships, we present a structured heterogeneous information network (HIN) \cite{sun2012mining} and use meta-path based approach \cite{Yizhou11} to build up relatednesses over apps. To efficiently classify nodes (e.g., apps) in the constructed HIN, HIN embedding methods \cite{fu2017hin2vec,dong2017metapath2vec,KDD2018} have been proposed. However, most of these existing methods are primarily designed for static networks, where all nodes are known before learning. As our application requires real-time prediction of new coming nodes (i.e., unknown apps) to detect Android malware, it is infeasible to rerun HIN embeddings whenever new nodes arrive, especially considering the fact that rerunning HIN embeddings also results in the need of retraining the downstream classifier. How to efficiently learn the representations of out-of-sample nodes in HIN, i.e. nodes that arrive after the HIN embedding process, remains largely unanswered. To solve this problem, we propose the \textit{HinLearning} method to first obtain in-sample node embeddings and then learn the representations of out-of-sample nodes in HIN, which is capable to preserve the heterogeneous property of HIN and also allows fast and scalable learning. Afterwards, we design a deep neural network (DNN) classifier leveraging the advantages of convolutional neural networks (CNNs) and Inception for Android malware detection. We develop a system  \textit{AiDroid} for real-time Android malware detection, which has the following major traits:

\vspace{-0.12cm}
\begin{itemize}
\item Besides runtime behaviors extracted from Android apps, we further analyze the complex relationships within the ecosystem (i.e., \textit{app-API}, \textit{app-IMEI}, \textit{app-signature}, \textit{app-affiliation}, \textit{IMEI-signature}, \textit{IMEI-affiliation} relations) to characterize Android apps. We then present HIN to represent the Android apps and exploit meta-path based approach to depict the relatednesses over apps. This provides a comprehensive solution that is capable to be more resilient against Android malware's evasion tactics.

\vspace{-0.12cm}	
\item We are the first to propose the method (denoted as \textit{HinLearning}) to efficiently learn the representations for out-of-sample nodes in HIN using in-sample node embeddings while without rerunning/adjusting them, which makes the downstream classifier feasible for classifying new arriving nodes (e.g., apps) without retraining. Though it's proposed for real-time Android malware detection, the \textit{HinLearning} method is a general framework which is able to learn desirable node representations in HIN (i.e., including in-sample and out-of-sample nodes) and thus can be further applied to various speed-sensitive dynamic network mining tasks (e.g., node classification, clustering).

\vspace{-0.12cm}	
\item We provide a comprehensive experimental study based on the large-scale real sample collections from Tencent Security Lab, which demonstrates the effectiveness and efficiency of our developed system \textit{AiDroid}. It has already been incorporated into Tencent Mobile Security product that server millions of users worldwide. 	
\end{itemize}

\section{2. System Overview}\label{systemArt}

The overview of our developed system \textit{AiDroid} for real-time Android malware detection is shown in Figure \ref{fig:system}, mainly consisting of the following components:

\vspace{-0.15cm}
\begin{itemize}

\item \textbf{Feature Extractor.} Through the installed Tencent Mobile Security product, smart phone users can upload the extracted API call sequences of Android apps as well as the meta-data to \textit{AiDroid}. It will then analyze various relationships among different types of entities (i.e., \textit{app}, \textit{API}, \textit{IMEI}, \textit{signature}, \textit{affiliation}) to depict the Android apps. (See Section 3.1. for details.) 

\vspace{-0.12cm}
\item \textbf{HIN Constructor.} In this module, based on the features extracted from the previous component, a structural HIN is first presented to model the relationships among different types of entities; and then different meta-paths are built from the HIN to capture the relatedness over apps from different views (i.e., with different semantic meanings). (See Section 3.2. for details.)

\vspace{-0.12cm}
\item \textbf{HIN Representation Learner.} In \textit{HinLearning}, based on the meta-path schemes built from the previous module, a heterogeneous in-sample node embedding method HINE is first proposed to learn the low-dimensional representations for in-sample nodes in HIN; then, Hin2Img method is devised to learn representations of out-of-sample nodes using learned in-sample node embeddings, which is also capable to enrich representations of in-sample nodes. (See Section 3.3. for details.)

\item \textbf{DNN Classifier.} After representation learning using \textit{HinLearning}, the learned representations of in-sample nodes with type of app will be fed to the designed DNN (see Section 3.4. for details) to train the classification model, based on which new arriving nodes (i.e., out-of-sample nodes with type of app) can be predicted as either benign or malicious. 
\end{itemize}

\section{3. Proposed Method} \label{method}
In this section, we introduce the detailed approaches of how we represent the Android apps, and how to solve  Android malware detection (i.e., classification) problem based on the representations.

\subsection{3.1. Feature Extraction} \label{FeatureExtraction}

\noindent \textbf{Dynamic Behavior Extraction.} API calls are used by Android apps in order to access Android OS functionality and system resources. Therefore, we extract the sequences of API calls in the application framework from runtime executions of Android apps to capture their behaviors. For example, a sequence of API calls (\textit{StartActivity}, \textit{checkConnect}, \textit{getPhoneInfo}, \textit{receiveMsg}, \textit{sendMsg}, \textit{finishActivity}) extracted from the previous mentioned ``TigerEyeing'' trojan represents its typical behaviors of connecting to the C\&C server in order to fetch the configuration information; while another sequence of its extracted API calls (\textit{startActivity}, \textit{checkConnect}, \textit{sendSMS}, \textit{finishActivity}) denotes its intention of sending SMS messages without user's concern. Note that, to simplify the further computation, each API call in the extracted sequences will be mapped to an integer ID.

\noindent \textbf{Relation-based Feature Extraction.} Besides the API call sequences extracted from an Android app that can be used to represent its behaviors, to detect the increasingly sophisticated Android malware, we further consider the following kinds of relationships. 

\vspace{-0.15cm}
\begin{itemize}

\item \textbf{\textit{R1}}: To describe the relation between an app and an API call it invokes during runtime execution, we build the \textbf{\textit{app-invoke-API}} matrix ${\bf I}$ where each element $i_{i,j} \in \{0,1\} $ denotes if app \textit{i} invokes API call \textit{j}.

\vspace{-0.12cm}
\item \textbf{\textit{R2}}: To describe the relation that an app exists (i.e., is installed) in a smart phone (i.e., IMEI), we generate the \textbf{\textit{app-exist-IMEI}} matrix ${\bf E}$ where each element $ e_{i,j} \in \{0,1\} $ means if app \textit{i} exists in phone \textit{j}.

\vspace{-0.12cm}	
\item \textbf{\textit{R3}}: Every app run on the Android platform must be signed by the developer. To depict such relationship, we build the \textbf{\textit{app-certify-signature}} matrix ${\bf C}$ whose element $ c_{i,j} \in \{0,1\} $ denotes if app \textit{i} is certified by signature \textit{j}.

\vspace{-0.12cm}	
\item \textbf{\textit{R4}}: Package name (a.k.a. Google Play ID) is the unique identifier for an Android app. Companies conventionally use their reversed domain name to begin their package names (e.g., ``com.tencent.mobileqq''). We extract the domain name from the package name to denote the relation between an app (e.g., ``mobileqq'') and its affiliation (e.g., ``tencent.com''); and then we generate the \textbf{\textit{app-associate-affiliation}} matrix ${\bf A}$ where each element $ a_{i,j} \in \{0,1\} $ indicates whether app \textit{i} is associated with affiliation \textit{j}.

\vspace{-0.12cm}
\item \textbf{\textit{R5}}: To represent the relation that a smart phone has a set of apps signed by a particular developer, we create the \textbf{\textit{IMEI-have-signature}} matrix ${\bf H}$ where each element $h_{i,j} \in \{0,1\} $ denotes if smart phone \textit{i} has signature \textit{j}.

\vspace{-0.12cm}	
\item \textbf{\textit{R6}}: To denote the relation that a smart phone installs a set of apps associated with a specific affiliation, we generate the \textbf{\textit{IMEI-possess-affiliation}} matrix ${\bf P}$ where each element $ p_{i,j} \in \{0,1\} $ denotes whether smart phone \textit{i} possesses affiliation \textit{j}.

\end{itemize}

\subsection{3.2. HIN Construction} \label{subsec:hin}

In order to depict apps, APIs, IMEIs, signatures and affiliations as well as the rich relationships among them (i.e., \textit{R1}-\textit{R6}), it is important to model them in a proper way so that different kinds of relations can be better and easier handled. We introduce how to use HIN, which is capable to be composed of different types of entities and relations, to represent the apps by using the features extracted above. We first present the concepts related to HIN as follows.

\vspace{-0.12cm}
\begin{definition}
A \textbf{\textit{heterogeneous information network (HIN)}} \cite{sun2012mining} is defined as a graph ${\mathcal G} = ({\mathcal V}, {\mathcal E})$ with an entity type mapping $\phi$: ${\mathcal V} \to \mathcal A$ and a relation type mapping $\psi$: ${\mathcal E} \to \mathcal R$, where ${\mathcal V}$ denotes the entity set and ${\mathcal E}$ is the relation set, $\mathcal A$ denotes the entity type set and $\mathcal R$ is the relation type set, and the number of entity types $|\mathcal A|>1$ or the number of relation types $|\mathcal R|>1$. The \textit{\textbf{network schema}} \cite{sun2012mining} for a HIN $\mathcal G$, denoted as $\mathcal T_{\mathcal G} = (\mathcal A, \mathcal R)$, is a graph with nodes as entity types from $\mathcal A$ and edges as relation types from $\mathcal R$.
\end{definition}
\vspace{-0.12cm}

\begin{figure}[htbp!]
	\vspace{-0.2cm}
	\centering
	\includegraphics[width=0.85\linewidth]{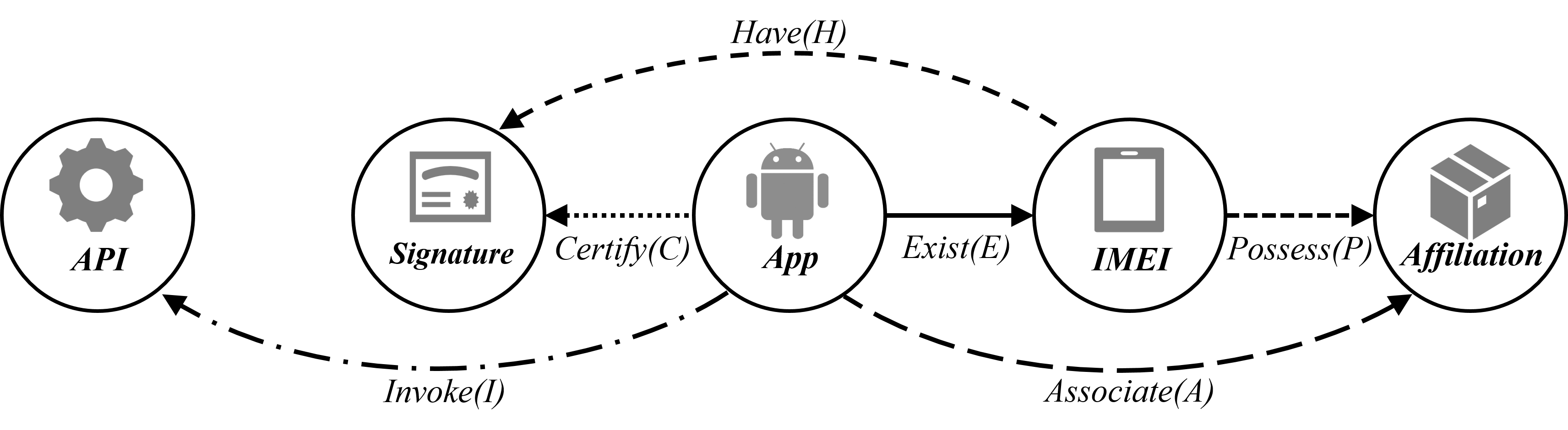}
	\vspace{-0.3cm}
	\caption{Network schema for HIN in our application.} \label{fig:networkschema}
	\vspace{-0.3cm}
\end{figure}

HIN not only provides the network structure of the data associations, but also provides a high-level abstraction of the categorical association. For our case, we have five entity types (i.e., app, API, IMEI, package, signature) and six types of relations among them (i.e., \textit{R1}-\textit{R6}). Based on the definitions above, the network schema for HIN in our application is shown in Figure \ref{fig:networkschema}, which enables the apps to be represented in a comprehensive way that utilizes their semantic and structural information.

\begin{figure}[htbp!]
	\vspace{-0.2cm}
	\centering
	\includegraphics[width=0.85\linewidth]{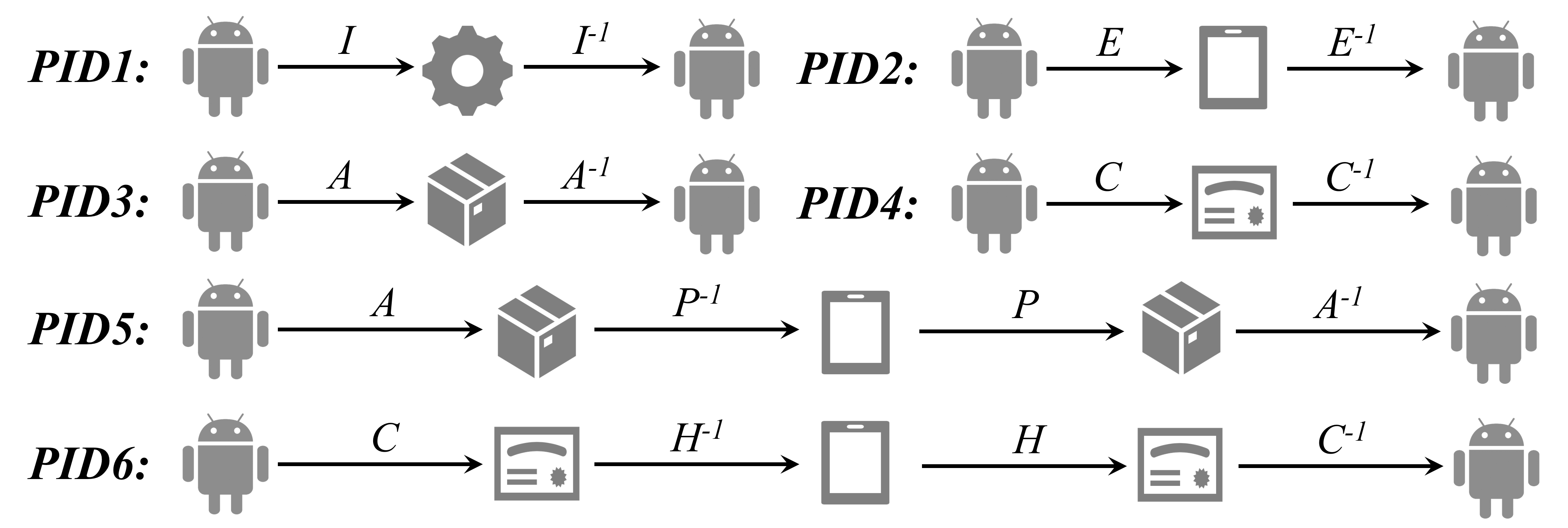}
	\vspace{-0.2cm}
	\caption{Meta-paths built for Android malware detection (the symbols are the abbreviations shown in Figure~\ref{fig:networkschema}).} \label{fig:metapaths}
	\vspace{-0.3cm}
\end{figure}

To formulate the relatedness among entities in HIN, the concept of meta-path has been proposed \cite{Yizhou11}: a \textbf{\textit{meta-path}} $ \mathcal{P} $ is a path defined on the graph of network schema $\mathcal T_{\mathcal G}=(\mathcal{A, R}) $, and is denoted in the form of $ A_{1} \xrightarrow{R_{1}} A_{2} \xrightarrow{R_{2}} ... \xrightarrow{R_{L}} A_{L+1} $, which defines a composite relation $R = R_1 \cdot R_2 \cdot \ldots \cdot R_L$ between types $A_1$ and $A_{L+1}$, where $\cdot$ denotes relation composition operator, and $L$ is length of $\mathcal P$. In our application, based on the HIN schema shown in Figure
\ref{fig:networkschema}, incorporated the domain knowledge from anti-malware experts, we design six meaningful  meta-paths to characterize the relatedness over apps at different views (i.e., \textit{PID1-PID6} shown in Figure \ref{fig:metapaths}).

For example, \textit{PID1} depicts that two apps are related if they both invoke the same API (e.g., two malicious mobile video players both invoke the API of ``\textit{requestAudioFocus}''); while \textit{PID5} describes that two apps are connected if their associated affiliations are possessed by (i.e., co-occur in) the same phone. To measure the relatedness over HIN entities (e.g., apps), traditional representation learning for HIN \cite{Yizhou11} mainly focuses on factorizing the matrix (e.g., adjacency matrix) of a HIN to generate latent-dimension features for the nodes in this HIN. However, the computational cost of decomposing a large-scale matrix is usually very expensive, and also suffers from its statistical performance drawback \cite{grover2016node2vec}. Since Android malware detection is a speed sensitive application and requires cost-effective solutions, scalable representation learning method for HIN, especially for out-of-sample nodes, is in need.

\subsection{3.3. \textit{HinLearning}: Representation Learning of In-Sample and Out-of-Sample Nodes in HIN} \label{subsec:detection}

To address the above challenge, we first formalize the problem of HIN representation learning as follow.

\vspace{-0.12cm}
\begin{definition}
	\textbf{\textit{HIN Representation Learning}} \cite{fu2017hin2vec,dong2017metapath2vec}. Given a HIN ${\mathcal G} = ({\mathcal V}, {\mathcal E})$, the representation learning task is to learn a function $f: \mathcal{V} \rightarrow \mathbb{R}^d $ that maps each node $v \in \mathcal{V}$ to a vector in a \textit{d}-dimensional space $ \mathbb{R}^d $, $d\ll |\mathcal V| $ that are capable to preserve the structural and semantic relations among them.
\end{definition}
\vspace{-0.12cm}

To solve the problem of HIN representation learning, due to the heterogeneous property of HIN (i.e., network consisting of multi-typed entities and relations), it is difficult to directly apply the conventional homogeneous network embedding techniques (e.g., DeepWalk \cite{perozzi2014deepwalk}, LINE \cite{tang2015line}, node2vec \cite{grover2016node2vec}) to learn the latent representations for HIN. To address this issue, HIN embedding methods \cite{fu2017hin2vec,dong2017metapath2vec,KDD2018} have been proposed, which are capable to preserve the semantic and structural correlations between different types of nodes . However, most of these existing methods are primarily designed for static networks, where all nodes are known before learning. In our application (i.e., real-time Android malware detection), it is infeasible to rerun HIN embeddings whenever new nodes arrive, especially considering the fact that rerunning HIN embeddings also results in the need of retraining the downstream classifier. How to efficiently learn the representations of out-of-sample nodes in HIN, i.e. nodes that arrive after the HIN embedding process, remains largely unanswered. To solve this problem, we first propose heterogeneous in-sample node embedding (HINE) model to learn in-sample node embeddings that is able to preserve the heterogeneous property of HIN; then, we devise Hin2Img to learn out-of-sample node representations and also enrich in-sample node representations using previous learned in-sample node embeddings without rerunning/adjusting HIN embeddings.

\noindent \textbf{HINE: Heterogeneous In-sample Node Embedding.} We first propose a random walk strategy guided by different meta-paths to map the word-context concept in a text corpus into a HIN; then we exploit skip-gram to learn effective in-sample node representations for a HIN.

Given a source node $v_j$ in a homogeneous network, the traditional random walk is a stochastic process with random variables $v_j^1, v_j^2,...,v_j^k$ such that $v_j^{k+1}$ is a node chosen at random from the neighbors of node $v_k$. The transition probability $p(v_j^{i+1}|v_j^i)$ at step $i$ is the normalized probability distributed over the neighbors of $ v_j^i $ by ignoring their node types. However, this mechanism is unable to capture the semantic and structural correlations among different types of nodes in a HIN. Here, we show how we use different built meta-paths to guide the random walker in a HIN to generate the paths of multiple types of nodes. Given a HIN ${\mathcal G} = ({\mathcal V}, {\mathcal E})$ with schema $\mathcal T_{\mathcal G} = (\mathcal A, \mathcal R)$, and a set of different meta-paths $\mathcal{S} = \{\mathcal{P}_{j}\}^{n}_{j=1}$, each of which is in the form of $A_{1} \rightarrow ... A_{t} \rightarrow A_{t+1} ... \rightarrow A_{l}$, we put a random walker to traverse the HIN. The random walker will first randomly choose a meta-path $\mathcal{P}_{k}$ from $\mathcal{S}$ and the transition probabilities at step \textit{i} are defined as:

\begin{equation}\label{eq:randomwalk}
\small
\begin{split}
&p(v^{i+1}|v_{A_{t}}^{i}, \mathcal{S}) = \\
&\begin{cases}
\frac{\lambda}{|\mathcal{S}|}\frac{1}{|N_{A_{t+1}}(v_{A_{t}}^{i})|}  \\ 
\ \ \ \ \ \ \ \ \ \mathrm{if}\ (v^{i+1},v^{i}_{A_{t}}) \in \mathcal{E},\phi(v^{i}_{A_{t}}) = A_{app}, \phi(v^{i+1}) = A_{t+1}\\ 
\frac{1}{|N_{A_{t+1}}(v_{A_{t}}^{i})|}   \\ 
\qquad\qquad\quad \qquad\qquad\mathrm{if}\ (v^{i+1},v^{i}_{A_{t}}) \in \mathcal{E},\phi(v^{i}_{A_{t}}) \ne A_{app},   \\
\qquad\qquad\quad\qquad \ \ \ \ \ \ \  \phi(v^{i+1}) = A_{t+1}, (A_{t},A_{t+1}) \in \mathcal{P}_{k} \\ 
0  \qquad\qquad\quad \qquad\qquad\quad \qquad\quad \qquad \qquad\ \ \ \otherwise,  
\end{cases}
\end{split}
\end{equation}

where $\phi$ is the node type mapping function, $N_{A_{t+1}}(v_{A_{t}}^{i})$ denotes the $A_{t+1}$ type of neighborhood of node $v_{A_{t}}^{i}$, $A_{app}$ is entity type of \textit{app}, and $\lambda$ is the number of meta-paths starting with $A_{app} \rightarrow A_{t+1}$ in the given meta-path set $\mathcal{S}$. The walk paths generated by the above strategy are able to preserve both the semantic and structural relations between different types of nodes in the HIN. 

After mapping the word-context concept in a text corpus into a HIN via the above proposed meta-path guided random walk strategy (i.e., a sentence in the corpus corresponds to a sampled path and a word corresponds to a node), skip-gram \cite{word2vec,perozzi2014deepwalk} is then applied on the paths to minimize the loss of observing a node's neighbourhood (within a window \textit{w}) conditioned on its current representation. The objective function of skip-gram is:

\begin{equation}\label{eq:log}
\arg\min_{X}\sum_{-w\leq k\leq w, j\neq k} -\log p(v_{j+k}|X(v_j)),
\end{equation}

where $X(v_j)$ is the current representation vector of $v_j$, and $p(v_{j+k}|X(v_j))$ is defined using the softmax function:
\begin{equation}\label{eq:softmax}
p(v_{j+k}|X(v_j))=\dfrac{\exp(X(v_{j+k})\cdot X(v_j))}{\sum_{q=1}^{|\mathcal{V}|} \exp(X(v_q)\cdot X(v_j))}.
\end{equation}
Due to its efficiency, we first apply hierarchical softmax technique \cite{mikolov2013distributed} to solve Eq. \ref{eq:softmax}, and employ the stochastic gradient descent to train the skip-gram.

\noindent \textbf{Hin2Img: Out-of-sample node representation learning.} Can we use in-sample node embeddings learned by the above proposed HINE to efficiently learn representations of out-of-sample nodes in HIN and also enrich in-sample node representations? To answer this question, we first present the concept of $k$-order neighbors in HIN as following.

\begin{definition}
	\textbf{\textit{k-order Neighbors in HIN}}. Given a HIN  ${\mathcal G} = ({\mathcal V}, {\mathcal E})$, let 1-order neighbors of a node $v_{i} \in {\mathcal V}$ be $S^{(1)}(v_{i})$ so that $S^{(1)}(v_{i}) = \{v_{j}|(v_{i}, v_{j}) \in {\mathcal E}\}$; then, $k$-order neighbors $S^{(k)}(v_{i})$ of a node $v_{i}$ ($k > 1$) can be denoted as $S^{(k)}(v_{i}) = \{S^{(1)}(v_{z})\setminus S^{(k-2)}(v_{i}), v_{z} \in S^{(k-1)}(v_{i})\}$. 
\end{definition}

\begin{figure}[htbp!]
	\vspace{-0.1cm}
	\centering
	\includegraphics[width=\linewidth]{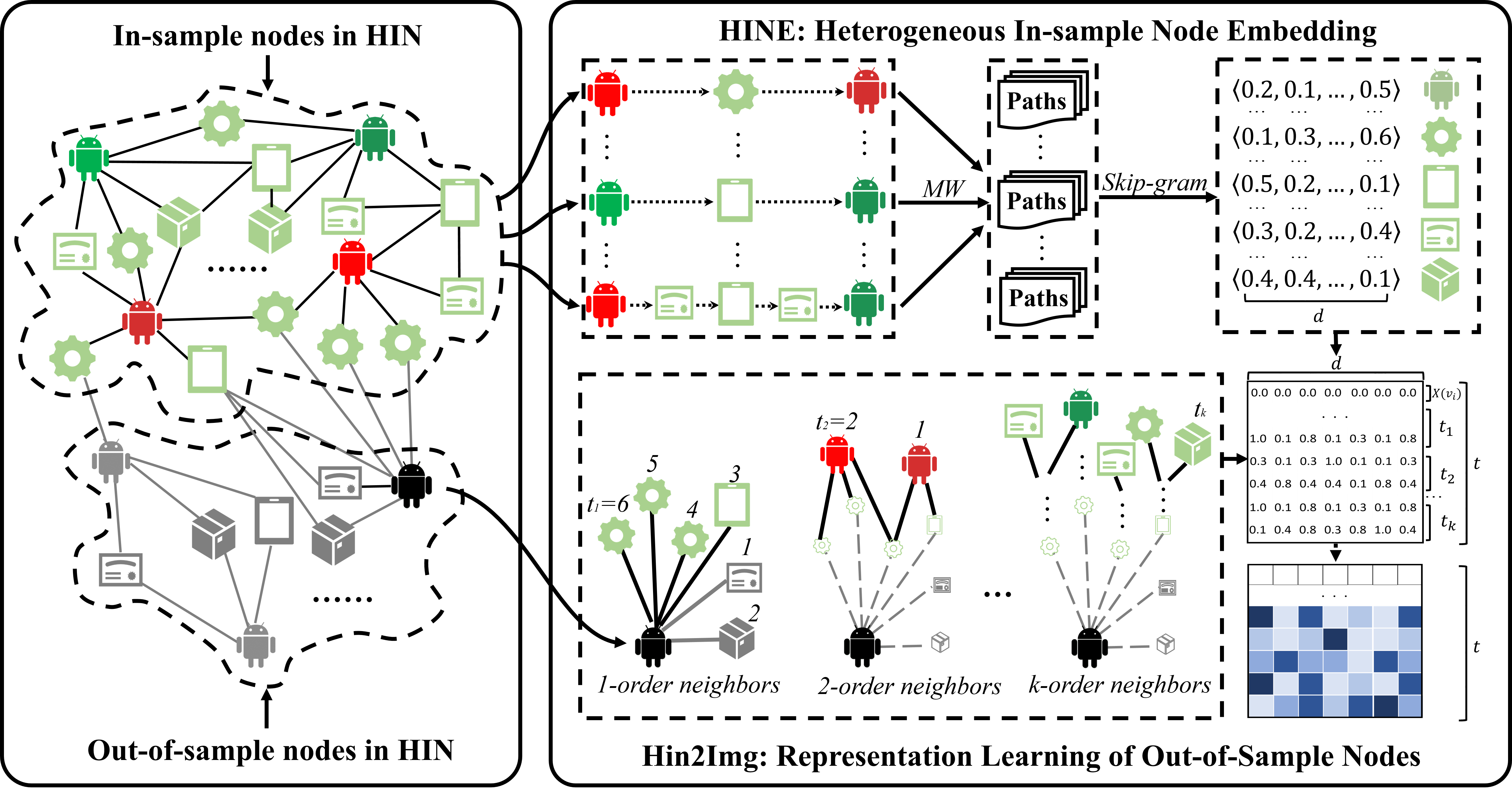}
	\vspace{-0.3cm}
	\caption{\textit{HinLearning}: node representation learning in HIN.} \label{fig:testing}
	\vspace{-0.3cm}
\end{figure}

By the above definition, each node (i.e., either in-sample or out-of-sample) in HIN is represented by the following representation matrix:

\begin{equation}\label{eq:X}
\mathbf{X}(v_{i}) = [X(v_{i}), X(S^{(1)}(v_{i})),..., X(S^{(k)}(v_{i}))],
\end{equation}

where $\mathbf{X} \in \mathbb{R}^{t \times d}$, each of which denotes $d$-dimensional node embedding. In our application, as illustrated in Figure \ref{fig:testing}, for each $k$, we exploit breadth-first search (BFS) method to find the $t_{k}$ neighbors in the order of type app, signature, package, IMEI and API, and $t = 1 + \sum_{1}^{k} t_{k}$. Empirically we found $k = 2$ and $t = d$ perform the best in our application, and apply them to our problem throughout the paper. Note that we zero-pad the representation matrix in the corresponding rows when the node embeddings cannot be found (i.e., out-of-sample nodes) or $t < d$.

Based on the in-sample node embeddings which can be learned offline using the proposed HINE, when a new node (e.g., a testing app) arrives, it only take $O(t \times d)$ time to obtain its representation using the proposed Hin2Img; furthermore, the representation learning for the new arriving node doesn't require rerunning HIN embeddings, which makes the downstream classifier workable for classifying the new arriving node without retraining.

\subsection{3.4. Deep Neural Network}

\begin{figure}[htbp!]
	\vspace{-0.2cm}
	\centering
	\includegraphics[width=\linewidth]{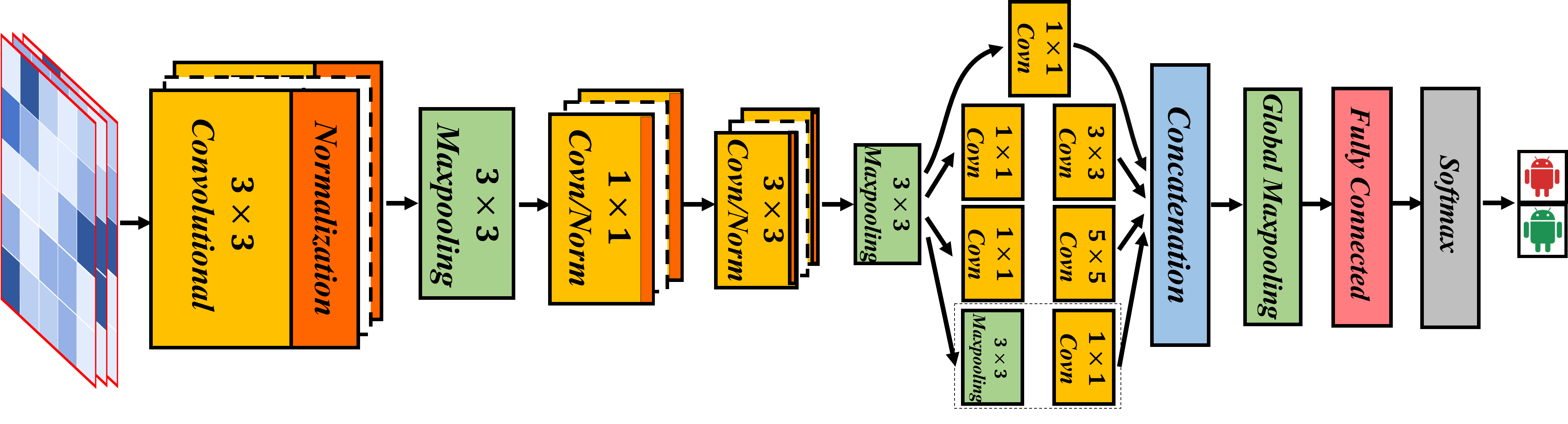}
	\vspace{-0.3cm}
	\caption{Our designed DNN for malware detection.} \label{fig:hincnn}
	\vspace{-0.3cm}
\end{figure}

CNNs \cite{LeCun_Nature_2015} have achieved great success in learning salient features for classification tasks; while the crafty architecture of Inception \cite{szegedy2015going} has shown high performance and low computational cost under strict constraints on memory and computational budget. Therefore, taking the generated representations of Android apps from previous section as inputs, we devise our deep learning framework leveraging the advantages of CNNs and Inception for real-time detection of Android malware, which is illustrated in Figure~\ref{fig:hincnn}. In our designed DNN, for training, we take the generated $t \times d$ representation matrix $\mathbf{X}$ of each in-sample node (with type of app) in HIN as input fed to the multilayer architecture to learn the higher level concept. In this multilayer architecture, the designed DNN first stacks pairs of convolutional layers and normalization layers with different filter sizes and strides followed by maxpoolings to capture universal features, such as curves and edges \cite{huang2018cost}. It then comes with an Inception module \cite{szegedy2015going} to generate more task-specific features, such as discriminative properties for malicious and benign apps. In the Inception module, the general features will be passed through the mixture of convolutional layers in parallel to take advantage of multi-level feature extraction, in which $1\times1$ convolutions and $3\times3$ maxpooling are configured for dimensionality reduction prior to $1\times1$, $3\times3$, and $5\times5$ convolutions that are used for feature learning, followed by a concatenation layer to concatenate the resulting feature representations. After the global maxpooling, based on a pair of fully connected layer and a softmax layer, the designed DNN will train the model for classifying any new arriving node (i.e., out-of-sample node with type of app represented as a $t \times d$ matrix $\mathbf{X}$) as either benign or malicious.

Algorithm~\ref{alg:algorithm} illustrates the implementation of our developed system \textit{AiDroid} which integrates the above proposed method in detail. 

\begin{algorithm}[htbp]
	\SetAlgoLined
	\KwIn{HIN ${\mathcal G} = ({\mathcal V}, {\mathcal E})$ with schema $\mathcal T_{\mathcal G} = (\mathcal A, \mathcal R)$, traning data set $D_{I}$ (in-sample apps), and testing data set $D_{O}$ (out-of-sample apps).}
	\KwOut{$\mathbf{y}$: The labels for the testing data set.}
	\BlankLine
	Learn in-sample node embeddings $X(v_{i}) \in \mathbb{R}^d$ ($i = 1, ..., |{\mathcal V}|$) using HINE\; 
	\For{$i = 1 \rightarrow |D_{I}|$}
	{
		Apply Hin2Img: for each $k$, find $k$-order neighbors $X(S^{(k)}(v_{i}))$ and then generate the $t \times d$ matrix $\mathbf{X}(v_{i}) = [X(v_{i}), X(S^{(1)}(v_{i})),..., X(S^{(k)}(v_{i}))]$\;		
	}
	Train DNN using $\mathbf{X}$s $\in \mathbb{R}^{t \times d}$\;
	\For{$j = 1 \rightarrow |D_{O}|$}
	{
		Use Hin2Img to generate $\mathbf{X}(v_{j})$\; 
		Obtain the label $y_{v_{j}}$ using trained DNN\;
	}
	return $\mathbf{y}$\;
	\caption{\textit{AiDroid}: HIN marries DNN.} \label{alg:algorithm}
\end{algorithm}

\section{4. Experimental Results and Analysis} \label{experiment}

In this section, we fully evaluate the performance of our developed system \textit{AiDroid} for Android malware detection.

\vspace{-0.2cm}
\subsection{4.1. Data Collection}

We obtain the large-scale real sample collection from Tencent Security Lab, which contains 190,696 training app (i.e., 83,784 benign and 106,912 malicious). After feature extraction and based on the designed network schema, the constructed HIN has 286,421 nodes (i.e., 190,696 nodes with type of app, 331 nodes with type of API, 70,187 nodes with type of IMEI, 8,499 nodes with type of signature, and 16,708 with type of affiliation) and 4,170,047 edges including relations of \textit{R1}-\textit{R6}. The new coming 17,746 unknown apps are used as testing data (to obtain the ground truth, they are further analyzed by the anti-malware experts, 13,313 of which are labeled as benign and 4,433 are malicious).

\vspace{-0.2cm}
\subsection{4.2. Baseline Methods}

We validate the performance of our proposed method in \textit{AiDroid} for Android malware detection by comparisons with different groups of baseline methods.

First, based on the constructed HIN described above, we evaluate our proposed HIN representation learn method \textit{HinLearning} by comparisons with following baselines.

\noindent \textbf{In-sample node embedding.} We compare our proposed HINE with other network embedding methods including:
\vspace{-0.12cm}
\begin{itemize}
\item \textbf{DeepWalk} and \textbf{LINE}: For DeepWalk \cite{perozzi2014deepwalk} and LINE \cite{tang2015line}, we ignore the heterogeneous property of HIN and directly feed the HIN for embedding.   

\vspace{-0.12cm}	
\item \textbf{metapath2vec}: We use each meta-path scheme separately to guide random walks in metapath2vec \cite{dong2017metapath2vec}.
\end{itemize}
\vspace{-0.12cm}

\noindent For HINE, we divide the designed meta-paths into three sets (i.e., $\mathcal{S}_{1}$ = $\{$\textit{PID1}$\}$, $\mathcal{S}_{2}$ = $\{$\textit{PID3, PID4}$\}$, $\mathcal{S}_{3}$ = $\{$\textit{PID2, PID5, PID6}$\}$) and use the proposed strategy to guide random walks. The parameter settings used for HINE are in line with DeepWalk, LINE and metapath2vec, which are empirically set as: vector dimension $d = 64$ (LINE: 64 for each order (1st- and 2nd-order)), walks per node $r = 20$, walk length $l = 50$ and window size $w = 5$.

\noindent \textbf{Out-of-sample node representation learning.} We compare our proposed Hin2Img for out-of-sample node representation learning with following baselines:

\begin{itemize}
\vspace{-0.12cm}
\item \textbf{LocalAvg}: The out-of-sample nodes are represented by averaging embeddings of neighboring in-sample nodes.
	
\vspace{-0.12cm}
\item \textbf{LabelProp}: Label propagation proposed for multivariate regression problem can be used to learn the representations of out-of-sample nodes. As it has been demonstrated \cite{ma2018depthlgp} that the vanilla version of label propagation \cite{vanillaLP} outperforms others, we hence use it as a baseline.

\vspace{-0.12cm}
\item \textbf{Rerunning}: For comparisons, we also run a baseline by rerunning all node embeddings when new nodes arrive.
\end{itemize}
\vspace{-0.12cm}

Second, we evaluate different types of features for Android malware detection. Our proposed method is general for HINs. Thus, a natural baseline is to see whether the knowledge we add in should be represented as HIN instead of other features. Here we compare two types of features: behavioral sequences and HIN-based features.

\begin{itemize}
\vspace{-0.12cm}
\item \textbf{Behavioral Sequences (\textit{f-1})}: We devise three baselines based on the extracted API call sequences of Android apps: (1) we build support vector machine (SVM) classifier based on binary (i.e., if an API call is invoked by an app) feature vectors (i.e., Bin+SVM); (2) we exploit Long Short-term Memory (LSTM) \cite{Sutskever} for sequence modeling, based on which SVM classifier is then built (i.e., LSTM+SVM); and (3) we also train our proposed DNN based on the extracted API call sequences for evaluation (i.e., Seq+DNN).   
	
\vspace{-0.12cm}
\item \textbf{HIN-based Features (\textit{f-2})}: Based on the constructed HIN, the proposed \textit{HinLearning} is applied for representation learning (i.e., both in-sample and out-of-sample nodes), based on which the designed DNN is used to train the classification model for prediction (i.e., \textit{AiDroid}). 
\end{itemize}
\vspace{-0.12cm}

\vspace{-0.2cm}
\subsection{4.3. Comparisons and Analysis}

Based on the HIN constructed from the training data, using the developed DNN as downstream classifier and the new coming nodes (i.e., apps) for testing, from Table~\ref{table:comparisons1}, we can observe that different combinations of in-sample node embedding and out-of-sample representation learning show different performances in Android malware detection: (1) For in-sample node embedding methods, our proposed HINE outperforms all baselines in terms of \textit{ACC} and \textit{F1}. That is to say, HINE learns significantly better node (i.e., app) representation in HIN than current state-of-the-art methods. The success of HINE lies in the proper consideration and accommodation of the heterogeneous property of HIN (i.e., the multiple types of nodes and relations), and the advantage of random walk guided by different meta-paths for sampling the node paths. (2) For out-of-sample representation learning, our proposed Hin2Img consistently and significantly outperforms all baselines (i.e., the detection performance achieve superb 0.9908 \textit{ACC} and 0.9817 \textit{F1}), which even surpasses the rerunning HIN embeddings. Obviously, $t \times d$ representation matrices learned by Hin2Img utilizing 1- and 2-order neighbors are more expressive than other embeddings in depicting the apps for the problem of real-time Android malware detection.

\vspace{-0.2cm}
\begin{table}[htbp]
	\centering
	\caption{Comparisons of different methods.}
	\vspace{-0.3cm}
	\label{table:comparisons1}
	\tabcolsep=1.5pt
	\begin{tabular}{cccccc}
		\toprule
		\multirow{2}{*}{{\small Metric}}&{\small In-sample}&\multicolumn{3}{c}{{\small Out-of-sample Learning}}&\multirow{2}{*}{{\small Rerunning}}\\
		\cmidrule(lr){3-5}
		& {\small Embedding}& {\small LocalAvg} & {\small LabelProp} & {\small \textbf{Hin2Img}} & \\
		\midrule
		\multirow{4}{*}{ACC}&DeepWalk& 0.9057& 0.9214& 0.9506&  0.9516\\
		& LINE & 0.9111 & 0.9307 & 0.9690 &  0.9602\\
		& metapath2vec & 0.9289 & 0.9448 & 0.9799 &  0.9722\\
		& \textbf{HINE} & 0.9389 & 0.9533 & \textbf{0.9908} &  0.9843\\
		\midrule
		\multirow{4}{*}{F1} &DeepWalk& 0.8267& 0.8547& 0.9055&  0.9076\\
		& LINE & 0.8364 & 0.8705 & 0.9398 &  0.9234\\
		& metapath2vec & 0.8669 & 0.8954 & 0.9606 &  0.9459\\
		& \textbf{HINE} & 0.8849 & 0.9094 & \textbf{0.9817} &  0.9691\\
		\bottomrule
	\end{tabular}
\end{table}
\vspace{-0.3cm}

We also show the comparisons in Table~\ref{table:comparisons2} for different features in Android malware detection. From the results, we can see that: (1) Based on the extracted API call sequences (i.e., \textit{f-1}), LSTM provides significant improvement in sequence modeling while our proposed DNN outperforms others in detection Android malware. (2) Compared with content-based features only, HIN-based features (i.e., \textit{f-2}) indeed perform better. The reason behind this is that HIN-based features are more expressive to characterize a complex and comprehensive relatednesses over apps through the designed meta-paths which consist of not only relationships between apps and their invoked API calls, but also higher-level semantics within the ecosystem.

\vspace{-0.3cm}
\begin{table}[htbp]
	\centering
	\caption{Comparisons of different types of features.}
	\vspace{-0.3cm}
	\label{table:comparisons2}
	\tabcolsep=2.5pt
	\begin{tabular}{cccccc}
		\toprule
		Feature&Method&TP&TN&FP&FN\\
		\midrule
		\multirow{3}{*}{\textit{\textbf{f-1}}}&Bin+SVM& 3,926& 11,828& 1,485 &  507\\
		& LSTM+SVM & 4,115 & 12,339 & 974 &  318\\
		& Seq+DNN & 4,168 & 12,504 & 809 &  265\\
		\textit{\textbf{f-2}} & \textbf{\textit{AiDroid}} & 4,395 & 13,188 & 125 &  38\\
		\midrule
		Feature&Method&Recall&Precision&ACC&F1\\
		\midrule
		\multirow{3}{*}{\textit{\textbf{f-1}}}&Bin+SVM& 0.8856& 0.7255& 0.8877&  0.7976\\
		& LSTM+SVM & 0.9282 & 0.8086 & 0.9271 &  0.8642\\
		& Seq+DNN & 0.9402 & 0.8374 & 0.9394 &  0.8858\\
		\textit{\textbf{f-2}} & \textbf{\textit{AiDroid}} & 0.9914 & 0.9723 & 0.9908 &  0.9817\\
		\bottomrule
	\end{tabular}
\end{table}
\vspace{-0.3cm}

\subsection{4.4. Parameter Sensitivity, Scalability and Stability}

\begin{figure}[htbp!]
	\vspace{-0.2cm}
	\centering
	\includegraphics[width=0.98\linewidth]{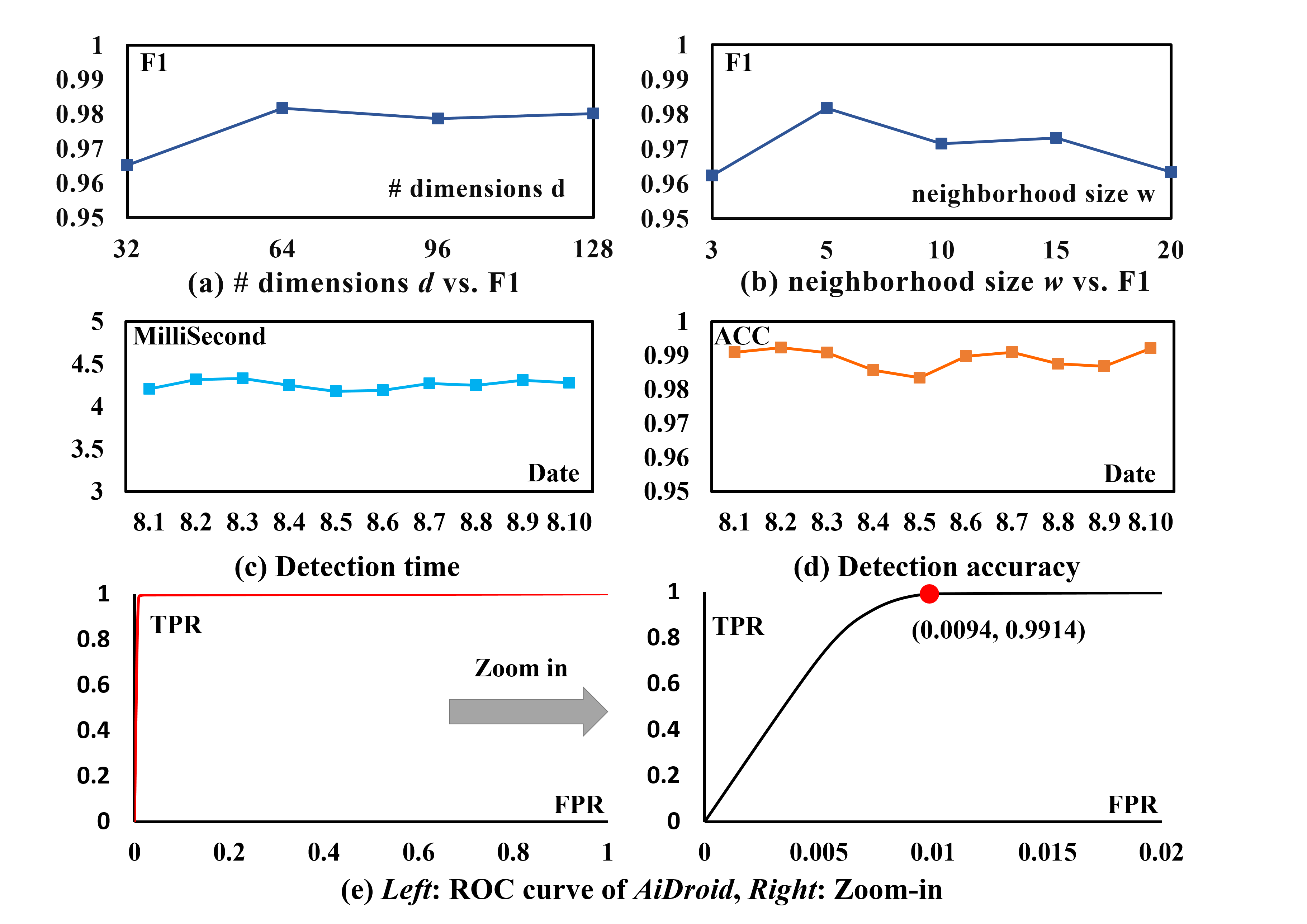}
	\vspace{-0.3cm}
	\caption{Parameter sensitivity, scalability and stability.} \label{fig:experiment}
	\vspace{-0.3cm}
\end{figure}

In this set of experiments, we systematically evaluate the parameter sensitivity, scalability and stability of our developed system \textit{AiDroid}. We first examine how latent dimensions (\textit{d}) and neighborhood size (\textit{w}) affect the performance in Android malware detection. As shown in Figure~\ref{fig:experiment}.(a) and (b), we can see that \textit{AiDroid} is not strictly sensitive to these parameters and is able to reach high performance under a cost-effective parameter choice. We then run the experiments using new arriving apps from Aug. 1-10, 2018 to assess the average detection time and accuracy. Figure~\ref{fig:experiment}.(c) and (d) demonstrate \textit{AiDroid} is scalable and stable over a long time span in detecting newly generated Android malware (i.e., average prediction time: 4.3 ms/app and 0.9891 ACC on average). Figure~\ref{fig:experiment}.(e) shows the ROC curve of \textit{AiDroid} based on the data described in Section 4.1 which achieves an impressive 0.9914 true positive rate (\textit{TPR}) at 0.0094 false positive rate (\textit{FPR}). We can conclude that \textit{AiDroid} is indeed feasible in practical use for real-time Android malware detection.

\section{5. Related Work} \label{relatedwork}

In recent years, there have been ample research studies on developing intelligent Android malware detection systems using machine learning and data mining techniques \cite{DroidMat,DroidDolphin,DroidDelver:2016,chen2017securedroid,Ye:CSUR,Madam_2018}. For example, DroidDolphin \cite{DroidDolphin} built classifiers based on dynamic analysis, while DroidMat \cite{DroidMat} and DroidMiner \cite{DroidMiner} constructed their models based on static analysis. However, most of the existing systems merely utilize content-based features for the detection. To further address the challenges of Android malware detection, in our preliminary work, HinDroid \cite{hou2017hindroid} was proposed which considered higher-level semantic relations among apps and APIs and introduced HIN for the first time in Android malware detection; but HinDroid was primarily designed for static HIN without considering new arriving nodes. 

To solve the problem of network representation learning, after DeepWalk \cite{perozzi2014deepwalk}, LINE \cite{tang2015line} and node2vec \cite{grover2016node2vec} that were proposed for homogeneous network embedding, HIN2vec \cite{fu2017hin2vec}, metapath2vec \cite{dong2017metapath2vec}, metagraph2vec \cite{KDD2018}, and PME \cite{PME_KDD2018} have been proposed for HIN representation learning. However, few of them can deal with out-of-sample nodes, i.e., nodes that arrive after the HIN embedding process. Though algorithms \cite{chang2015heterogeneous,zhao2018optimization} have been proposed to infer embeddings for out-of-sample nodes in HIN, they necessitate adjusting in-sample node embeddings and also the downstream classifier retraining. Efficient representation learning for out-of-sample nodes in HIN without rerunning/adjusting HIN embeddings is in need for our application in real-time Android malware detection.

\section{6. Conclusion} \label{conclusion}

To combat the evolving Android malware attacks, in this paper, we first extract the API call sequences from runtime executions of Android apps and further analyze higher-level semantic relationships within the ecosystem. To depict such complex relations, we introduce HIN for modeling and use meta-path based approach to build up relatednesses over apps. To efficiently classify nodes (i.e., apps) in HIN, we propose the \textit{HinLearning} method to first gain in-sample node embeddings and then learn representations of out-of-sample nodes without rerunning/adjusting HIN embeddings for the first time. Afterwards, we design a DNN classifier leveraging the advantages of CNNs and Inception for Android malware detection. A comprehensive experimental study on the large-scale real data collections from Tencent Security Lab is performed to compare various baselines. Promising experimental results demonstrate that our developed system \textit{AiDroid} outperforms others in real-time Android malware detection, which has been incorporated into Tencent Mobile Security product that serves millions of users worldwide. 

\section{Acknowledgement} \label{acknowledgement}

The authors would also like to thank the anti-malware experts of Tencent Security Lab (Yinming Mei, Yuanhai Luo, Hong Yi, and Kui Wang) for helpful discussion and implementation. Y. Ye, S. Hou, and L. Chen's work is partially supported by the U.S. National Science Foundation under grants CNS-1618629, CNS-1814825 and OAC-1839909, WV HEPC.dsr.18.5, and WVU Research and Scholarship Advancement Grant (R-844).

\bibliographystyle{aaai}
\bibliography{AiDroid_Ye}

\end{document}